\documentclass[conference]{IEEEtran}
\usepackage{cite}
\usepackage{amsmath}
\usepackage{algorithm}
\usepackage{algorithmic}
\usepackage{mathabx}
\usepackage{multicol}
\usepackage{subcaption}
\usepackage{color}
\usepackage{graphicx}
\usepackage{epstopdf}
\usepackage{color,soul}

\usepackage{multirow}

\begin{document}

\title{Joint User Association and BS Switching Scheme for Green Heterogeneous Cellular Network}

\author{\IEEEauthorblockN{Waqas ur Rehman}
\IEEEauthorblockA{School of Electrical Engineering\\
National University of Computer and\\
Emerging Science, Lahore, Pakistan\\
Email: waqas.rehman@nu.edu.pk}
\and
\IEEEauthorblockN{Arshad Hussain}
\IEEEauthorblockA{School of Engineering\\
Univeristy of Central Punjab\\
Lahore, Pakistan\\
Email: arshad.hussain@ucp.edu.pk}
\and
\IEEEauthorblockN{M. Majid Butt}
\IEEEauthorblockA{Nokia Bell Labs\\
Paris-Saclay, France\\
E-mail: majid.butt@ieee.org}}

\renewcommand\IEEEkeywordsname{Keywords}

\maketitle

\begin{abstract}
A dense deployment of on-grid small cell base stations (SBSs) in a heterogeneous cellular network (HCN) results in large power consumption which can be reduced by utilizing harvested energy from scavenge sources. Recently, a new layout for HCN with energy harvesting-SBSs (EH-SBSs) has been proposed to reduce inter-cell interference and power consumption; and increase system energy efficiency. In this paper, an energy efficient joint user association and BS on-off scheme for an HCN having diverse energy source(s) is proposed, which aims at utilizing the harvested energy efficiently in order to reduce the on-grid power consumption. The scheme operates in two phases. First, it decides which BS should be in off, sleep or active mode and iteratively updates its transmit power based on energy arrival and consumption, to minimize interference and on-grid power consumption. Secondly, for further reduction in on-grid power consumption, the user discovers and associates itself with the nearest available EH-SBS. The system energy consumption for the proposed scheme is numerically evaluated using Monte Carlo simulations. Simulation results reveal that the proposed scheme shows significantly better results as compared to other schemes in the literature in terms of grid power consumption and energy efficiency.
\end{abstract}


\IEEEpeerreviewmaketitle
\section{Introduction}
In order to increase spectral efficiency, heterogeneous networks have been introduced in LTE-Advanced where different types of base stations (BS) are considered instead of conventional macro BSs (MBSs). To fill the coverage holes in the conventional system, low power small cell base stations (SBSs) were introduced in order to increase system capacity. Dense deployment of MBSs and SBSs requires large power consumption and burdens both, the network operator and power grid.  To address the large energy consumption problem, energy harvesting from scavenged sources is a promising solution. SBSs require low power, hence devices that harvest energy through solar, wind and radio frequency (RF) can be installed on SBSs. Self-powered cellular network and system cost of EH-SBSs have been evaluated in \cite{piro2013hetnets}  and it has been suggested that energy harvesting is an economical alternative to the existing system. Different traffic offloading schemes have been proposed for conventional cellular network in \cite{zhang2015many} and \cite{zhang2014energy}, but these schemes are not applicable to heterogeneous cellular networks (HCNs) having energy harvesting-SBSs (EH-SBSs).

In HCN, MBSs always act as a blanket and cover a large area. However, SBSs are dynamically activated and deactivated on the basis of traffic and energy status. For example, if a conventional SBS is lightly loaded, it must be in off state to reduce the on-grid power consumption, while EH-SBS must be deactivated if it has insufficient backup energy for transmission. Energy-aware traffic offloading schemes for single tier homogeneous and two tier heterogeneous networks have been proposed in \cite{gong2014base} and \cite{feng2014energy}. Joint user association and BS switching on/off algorithm was proposed in \cite{kim2013joint}, to balance energy consumption and revenue in heterogeneous networks.

Several studies address BS selection and user association policies under different models and assumptions \cite{Energy-Awaretraffic, ye2013user, jo2012heterogeneous, muhammad2017cell}. They consider different metrics for selecting the serving BS, e.g., nearest BS, load balancing or flexible cell association, etc. The scheme proposed in this paper with flexible cell area, jointly uses BS on-off switching and user association, which is shown more energy efficient and offers more system capacity. The association schemes in \cite{zhang2014energy}, \cite{Energy-Awaretraffic} assume a constant BS transmit power, whereas in the proposed approach, the transmit power level is updated iteratively on the basis of energy arrival and consumption, to minimize on-grid power consumption.

In this paper, we propose an energy efficient joint user association and BS switching scheme that increases system capacity and reduces on-grid power consumption. The users in this scheme are associated with the nearest EH-SBS, where each EH-SBS iteratively updates its coverage area to balance the load with harvested energy. The proposed scheme decides which conventional and EH-SBS should be in active-sleep and active-off modes, respectively. Simulation results demonstrate that more than 70\% of the on-grid power consumption can be saved for low user density.

The rest of paper is organized as follows. A system model for HCN performance evaluation is described in section II. The proposed user association and BS switching scheme is discussed in Section III. We numerically evaluate the performance of the proposed scheme in Section IV and conclude the paper in Section V.

\section{System Model}
\begin{figure}
	\centering
	\includegraphics[width=3.2in]{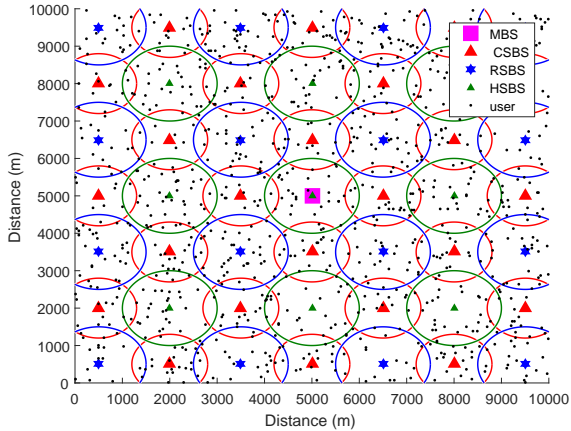}
	\caption{\small Illustration of an HCN with multiple CSBSs surrounding each MBS and multiple EH-SBSs surrounding each CSBSs. Only a single macro-cell is shown for the sake of simplicity.}
	\label{fig1}
\end{figure}

We consider an HCN with SBSs of different types in terms of energy supply, while MBS is connected with grid power as shown in Fig.~\ref{fig1}. SBSs with diverse sources can be classified into three different types:
\begin{itemize}
	\item Conventional small cell base stations (CSBSs): An SBS powered by on-grid power.
	\item Renewable small cell base stations (RSBSs): An EH-SBS powered by renewable energy sources (like solar and/or wind power).
	\item Hybrid small cell base stations (HSBSs): An EH-SBS powered by both on-grid and renewable energy sources.
\end{itemize}

RSBS operates with energy harvested from the environment and efficiently utilizes this energy to provide services to the associated users. Energy arrival is modeled using Poisson process with arrival rate $\lambda_E$. We use \textit{harvest-use} model where the harvested energy in a time slot is immediately used in the time slot and not stored \cite{krikidis2013harvest}.\footnote{Despite its simplicity as compared to harvest-store-use model, it provides valuable insight for system design.} Without loss of generality, the length of time slot is normalized to one, therefore energy and power terms can be used interchangeably. RSBS provides service(s) only if it has harvested enough energy in a time slot, otherwise it switches off, whereas an HSBS switches to on-grid mode.

Schematic diagram of the system layout is shown in Fig.~\ref{fig1}, where multiple CSBSs surround each MBS and similarly, multiple EH-SBSs (both RSBSs and HSBSs) surround each CSBS. For a small number of users, EH-SBSs manage the network load, whereas CSBSs are kept in sleep mode to minimize interference and the on grid power consumption. As demand for system capacity increases with an increase in the number of users, CSBSs are switched to active mode.

With respect to the location and BS association, a user is classified into two different types. If a user lies inside the coverage area of MBS (outside of the SBS) and is served by the MBS, it is classified as Macro-Macro User (MMU). Similarly, if a user lies inside the coverage area of SBS (which is in off/sleep mode) and is served by the other SBS or MBS, it is classified as Small-Small User (SSU) or Small-Macro User (SMU), respectively.

MBSs are grid powered base stations that have a large coverage area. These MBSs are always in active mode and can serve any user in the network. The user distribution in the spatial domain is modeled as a Poisson Point Process (PPP) with density $\rho$. As for spectrum resource, bandwidth available to the users is orthogonal to avoid inter cell interference, whereas spectrum reuse factor is 1.

\subsection{Power Consumption}
A base station can be in off, sleep or active mode. A user can only associate itself with an active BS. An active BS serving no user must be switched to sleep or off mode in order to reduce power consumption. For active mode, power consumption is defined in \cite{auer2010d2} as,
\begin{equation} \label{eq:one}
P_{BS}= P_c  + \beta \frac{w}{W}P_t \quad  \quad       0\leq w \leq W
\end{equation}
Here $\beta$ is the inverse of power amplifier efficiency, $P_c$ (constant power) represents the power required for baseband and system cooling. $W$ is the system's total bandwidth divided into $N$ orthogonal sub-carriers, $w$ is the bandwidth utilized by active sub-carriers and $P_t$ is the transmitted power. If no user is associated with a BS, the second term (RF power) in~\eqref{eq:one} becomes zero which shows that the BS still consumes some power to remain in active mode.

An RSBS can be in an active or off mode. In off mode, the BS is completely switched off and cannot provide service(s) to any user. Whereas, sleep mode is special type of off mode in which, BS does not serve any user and  transmits a beacon periodically to indicate its availability to the users. If the number of users a BS can serve is more than a threshold value $U_{min}$, the BS will switch to active mode. CSBS can only be in active or sleep mode.

An HSBS always remains in active mode and can provide services to more users in order to keep more CSBSs in sleep mode by utilizing both the EH and on-grid sources. The states of various types of BSs are summarized in TABLE~\ref{table:1}.
\begin{table}[h!]
		\renewcommand{\arraystretch}{1.0}
	\caption{Mode(s) adopted by SBS and MBS}
	\centering
	\begin{tabular}{lcccc}
		\hline
		\multirow{2}{1.5cm}{\textbf{Modes}}& \multicolumn{4}{c}{\centering \textbf{Base Stations}} \\
		\cline{2-5} & \multicolumn{1}{p{1.25cm}}{\centering MBS} & \multicolumn{1}{p{1.25cm}}{\centering CSBS} & \multicolumn{1}{p{1.25cm}}{\centering RSBS} &\multicolumn{1}{p{1.25cm}}{\centering HSBS} \\ \hline
		Active & Yes & Yes & Yes & Yes  \\
		Sleep & No & Yes & No  & No\\
		Off & No & No & Yes & No\\
		\hline
	\end{tabular}
\label{table:1}
\vspace{-0.4cm}
\end{table}

\subsection{Network Throughput}
In the downlink of HCN, network throughput is defined as the sum rate received by the users. The achievable rate $R$ for a user $i$ is given by
\begin{equation} \label{eq:two}
R_i = B \log_2 (1 + SINR_i)
\end{equation}
where $B$ is the fixed bandwidth allocated to a user. In HCN, generally the thermal noise is negligible compared to interference, therefore only the signal-to-interference ratio (SIR) is considered instead of signal-to-interference-plus noise ratio (SINR). As resource blocks are orthogonally allocated and frequency spectrum is reused, the SIR of the user at position $x$, served by $BS_y$, located at $y$, is given by \cite{Personalcell_Traffic_offloading},
\begin{equation} \label{eq:sir}
SIR_{y,x}=\frac{P_y h_{x,y} r_{x,y}^{-\alpha }}{\sum_{ z=\mathcal{B}\setminus y} P_z h_{x,z} r_{x,z}^{-\alpha}}
\end{equation}
A user can discover the set of BSs, $\mathcal{B}$ = \{$\mathcal{B}_R$, $\mathcal{B}_H$, $\mathcal{B}_C$, $\mathcal{B}_M$\} where $\mathcal{B}_R$, $\mathcal{B}_H$, $\mathcal{B}_C$ and $\mathcal{B}_M$ represent the set of RSBS, HSBS, CSBS and MBS, respectively to which a user can associate itself. If a user associates itself with $BS_y$, the set of BSs $\mathcal{B}\setminus y$, interferes with the associated user. $P_y$ is the transmit power of $BS_y$, $\alpha$ is the path loss exponent, $h_{x,y}$ is the channel gain and $r_{x,y} = ||x-y||$ is the Euclidean distance between a user located at $x$ and BS at $y$. Similarly, $h_{x,z}$ is the channel gain and $r_{x,z} = ||x-z||$ is the Euclidean distance between a user located at $x$ and any BS (not located at $y$) at $z$.
 
\subsection{Energy Efficiency}
Energy efficiency of a system is defined as ratio of sum rate (throughput) $R$ to on-grid power consumption $P$ (sum over all BS powers $P_{BS}$) in a network. Energy efficiency ($EE$) can be written as,
\begin{equation} \label{eq:two}
EE = \frac{R}{P}
\end{equation}
Energy efficiency can be optimized by enhancing the system capacity and reducing the on-grid power consumption. System capacity can be enhanced by moving CSBSs from the sleep mode to active mode at the additional cost of on-grid power.

\section{Power Consumption Minimization}
In this section, a novel joint BS on-off and user association scheme is presented to minimize the on-gird power consumption. Consider an MBS with $N_R$, $N_H$ and $N_C$ RSBS, HSBS and CSBS, respectively. Then, the total power consumption $P$ is given by,

	\begin{equation} \label{eq:P_sum}
	P = P_{M}+\sum_{j}^{N_R} I_{R_j} P_{R_j}+\sum_{k}^{N_H} P_{H_k}+ \sum_{i}^{N_C} I_{C_i} P_{C_i}
	\end{equation}
where $P_M$, $P_R$, $P_H$ and $P_C$ represent power consumed by MBS, RSBS, HSBS and CSBS respectively. $I_C$ and $I_R$ are indicator functions representing if CSBS and RSBS are in active-sleep and active-off mode, respectively.

The power consumption for MBS is expressed as,
\begin{equation}\label{eq:P_mbs}
\begin{split}
	P_{M} = P_{cm}+ \dfrac{P_{tm} \beta_m }{W_m} \left(w_{mmu} +  \sum_{i}^{N_R}(1- I_{R_i} ) w_{smu}^{r_i} \right. \\
\left. + \sum_{j}^{N_C} (1-I_{C_j})w_{smu}^{c_j} \right)
\end{split}
\end{equation}
where $w_{mmu}$, $w_{smu}^c$ and $w_{smu}^r$ represent the bandwidth used by MMUs and SMUs respectively, when CSBS(s) and RSBS(s) are in sleep and off mode. If all CSBSs and RSBSs are in an active state, then very few users are associated with MBS and RF power consumption is small. MBS provide services to the offloaded users when some of the CSBSs/RSBSs are in sleep/off state, and its RF power increases with the increase in number of associated users.

The power consumption for HSBS and CSBS is expressed as,
\begin{equation}\label{eq:P_hsbs}
\begin{split}
	P_{H}= P_{ch}  + \frac{\beta_c  P_{th}}{W_H}  \left( w_{ssu} +  \sum_{i}^{N_R} (1-I_{R_i})w_{ssu}^{r_i} \right. \\
\left. +\sum_{j}^{N_C} (1-I_{C_j})w_{ssu}^{c_j} \right) 
\end{split}
\end{equation}
\begin{equation} \label{eq:P_csbs}
P_{C}= P_{cc}  + \beta_c \frac{w_{ssu}}{W_C}  P_{TC}  
\end{equation}
Here, $w_{ssu}^c$ and $w_{ssu}^r$ represent the bandwidth used by SSUs, when CSBS and RSBS are in sleep and off mode respectively.

Sum rate of a system is the sum of all BSs' rates. Each BS rate is the sum of rates of the users associated with the particular BS. If the number of users associated with MBS, CSBS and RSBS are denoted by $U_M$, $U_C$ and $U_R$ respectively, and the bandwidth allocated to each user is fixed, the system capacity is given by,
\begin{align}
{R} &= \sum_{i}^{N_M} B\log_2 (1+SIR_i) +\sum_{l}^{N_{R}} \sum_{m}^{U_R} I_{R_l} B\log_2 (1+SIR_{l,m}) \nonumber \\
& +\sum_{n}^{N_{H}} \sum_{o}^{U_H} B\log_2 (1+SIR_{n,o}) \nonumber \\
& +\sum_{p}^{N_{C}} \sum_{q}^{U_R} I_{C_p} B\log_2 (1+SIR_{p,q})
\end{align}

System capacity is maximized when all BSs are in active state with additional on-grid power consumed by HSBSs and RSBSs. There is a trade-off between power consumption and throughput to maximize energy efficiency. To minimize the on-grid power consumption, a joint load balancing and cell association scheme is presented in next section.

\subsection{Joint User Association and BS on-off Scheduling}
The SBS with more harvested energy will try to serve more users by utilizing its green energy in an efficient way. To minimize power consumption, the user associates according to the following algorithm.\\
\textbf{User end}
\begin{itemize}
	\item First, the user discovers the number of EH-SBSs.\footnote{EH-SBSs means both RSBSs and HSBSs.} This	can be easily identified from the beacons which are periodically transmitted by the SBSs.
	\item A user than associates itself with a single EH-SBS based on certain parameter(s), such as nearest BS, maximum received power strength, maximum received SINR, etc. In order to improve the SIR and reduce transmission power, nearest EH-SBS association scheme is adopted.
	\item If a user cannot discover any RSBS, then nearest HSBS provides service(s) to the offloaded user.
	\item If a user does not discover any HSBS for association as well, then nearest CSBS serves the user.
	\item If no small cell serves the user, MBS provides service to the user.
	\end{itemize}
On the other side, EH-SBS changes its cell size by varying its transmit power based on load and harvested energy. Let the harvested energy in a time slot and energy consumption for RSBS transmit power $P_T$ be denoted by $E_{EH}$ and $E_{SBS}$, respectively, and $\Delta$ is defined as $E_{EH} - E_{SBS}$. Efficient use of the harvested energy for \textit{harvest-use} requires that all the harvested energy must be used for transmission. Due to practical reasons, we introduce a margin parameter $C >$ 0, which allows some margin between the harvested and consumed RSBS energy, thereby facilitating system design, i.e. $\Delta \geq$  C. When $E_{SBS} <  E_{EH}$ for power $P_{BS}$, the RSBSs reduce their transmit power to match the harvested energy.

To reduce inter-cell interference and balance the load with harvested energy, the base station works as follows: \\
\textbf{Base station end}
\begin{itemize}
	\item To indicate the transmit power  and the availability of harvested power, EH-SBSs periodically transmits a beacon signal to near CSBS and users respectively.
	\item When $\Delta >$ 0\\
	-- EH-SBSs increase their transmit power to provide service(s) to more offloaded users from other CSBSs. \\
	-- At the same time transmit power of CSBS is reduced to minimize interference and on-grid power consumption.\\
	-- CSBS switches to sleep mode (SBS periodically senses the channel), if the number of users associated are less than $U_{min}$.
	\item $\Delta \leq $ 0\\
	-- Transmit power of RSBS is reduced. CSBS is kept in sleep mode to reduce on-grid consumption. Transmit power of HSBS is increased to provide services to offloaded user. This is done by utilizing both harvested and on-grid energy.\\
	-- HSBS transmit power reaches maximum $P_{max}$ when the number of associated users reaches $N_{th}$. \\
	-- When HSBS is at $P_{max}$ while the number of users are increasing, EH-SBSs cannot provide service to all users. As a result, a few CSBSs switch into active mode that can provide service to more than $U_{min}$ users. If the number of users are still increasing, CSBS transmit power increases.\\
	-- If reducing transmit power of RSBS creates a coverage hole, MBS serves the remaining users. \\
	-- To further reduce on-gird power consumption, MBS provide service(s) to the users at its full capacity by switching some of CSBSs into sleep mode. We show in Section IV that this unconventional action helps to further reduce power consumption.
	\item Transmit power of SBS reaches a minimum value $P_{min}$, when the number of users associated are $U_{min}$. \\
	-- CSBS switches to sleep mode, while RSBS switches to off mode.
	\end{itemize}
The pseudo-code for joint user association and BS on/off scheme is summarized in Algorithm 1. This algorithm is run locally by every RSBS.

	\begin{algorithm}[ht!]
	\begin{algorithmic}[1]
		\STATE $\Delta$ = $E_{EH} - E_{SBS}$
		\WHILE{$\Delta \not\in [0,C]$ }
		\IF{$\Delta$ $\geq$ C }
		\STATE increase $P_T$ of $RSBS$
		\STATE increase $P_T$ of $HSBS$
		\STATE decrease $P_T$ of near $CSBSs$
		\IF{$CSBS$ users $\geq$ $U_{min} $ }
		\STATE switch CSBS in active mode
		\ELSE
		\STATE switch CSBS in sleep mode
		\ENDIF
		\ELSIF{$\Delta \leq $ 0}
		\STATE	decrease $P_T$ value of RSBS\\
		\IF{HSBS $P_T$ $\leq$ $P_{max}$ }
		\STATE increase $P_T$ of near HSBS
		\STATE	decrease $P_T$ of near CSBS
		\ELSE
		\STATE	increase $P_T$ of near CSBS
		\ENDIF
		\IF{RSBS $P_T$ $<$ $P_{min}$ }
		\STATE	switch RSBS in off mode
		\ELSE
		\STATE	switch RSBS in active mode
		\ENDIF
		\ENDIF
		\STATE $\Delta$ = $E_{EH} - E_{SBS}$
		\ENDWHILE	
		\WHILE{MBS users $\leq$ $N_{th}$}
		\STATE Find CSBS having minimum associated users
		\STATE Move CSBS into sleep mode and offload users to MBS
		\ENDWHILE
		\STATE Perform user association Scheme
	\end{algorithmic}
	\caption{New Joint Association Scheme}
\end{algorithm}
\section{Performance Evaluation}
In this section, evaluation of the proposed scheme is performed in terms of power consumption, throughput and energy efficiency. We evaluate the results using Monte Carlo simulations. The number of users and their locations are randomly generated, and the results are averaged over 5,000 simulation samples, while the margin parameter $C = 1$ for the simulations. For the simulation scenario, the parameter values in Table~\ref{table:two} are incorporated from \cite{tombaz2011impact}.
\begin{table}[ht]
	\caption{Simulation assumptions}
	\centering 
	\begin{tabular}{l c c} 
		\hline 
		\textbf{BS parameter} & \textbf{MBS} & \textbf{CSBS / RSBS / HSBS}\\[0.5ex]
		{Constant power $P_c$}  & 354.44 W & 38 W \\[0.5ex]
		{Transmit Power $P_t$} & 40 W & 30 W \\[0.5ex]
		{Coefficient $\beta$} & 21.45 & 5.5 \\[0.5ex]
		{Coverage area $D$} &  1730 m & 350 m \\[0.5ex]
		Path loss exponent $\alpha$ & 3.5 & 3.5 \\[0.5ex]
		Bandwidth $W$ & 10 MHz & 5 MHz  \\[0.5ex]
		Number of sub-carriers &1000 & 500 \\[0.5ex]
		Number of BS & 1 & 24 / 16 / 9
		\\[0.5ex]
		\hline 
	\end{tabular}
	\label{table:two} 
\end{table}
%
%
%

\begin{figure*}
	\begin{multicols}{4}
		\centering
		\includegraphics[width=1.9in]{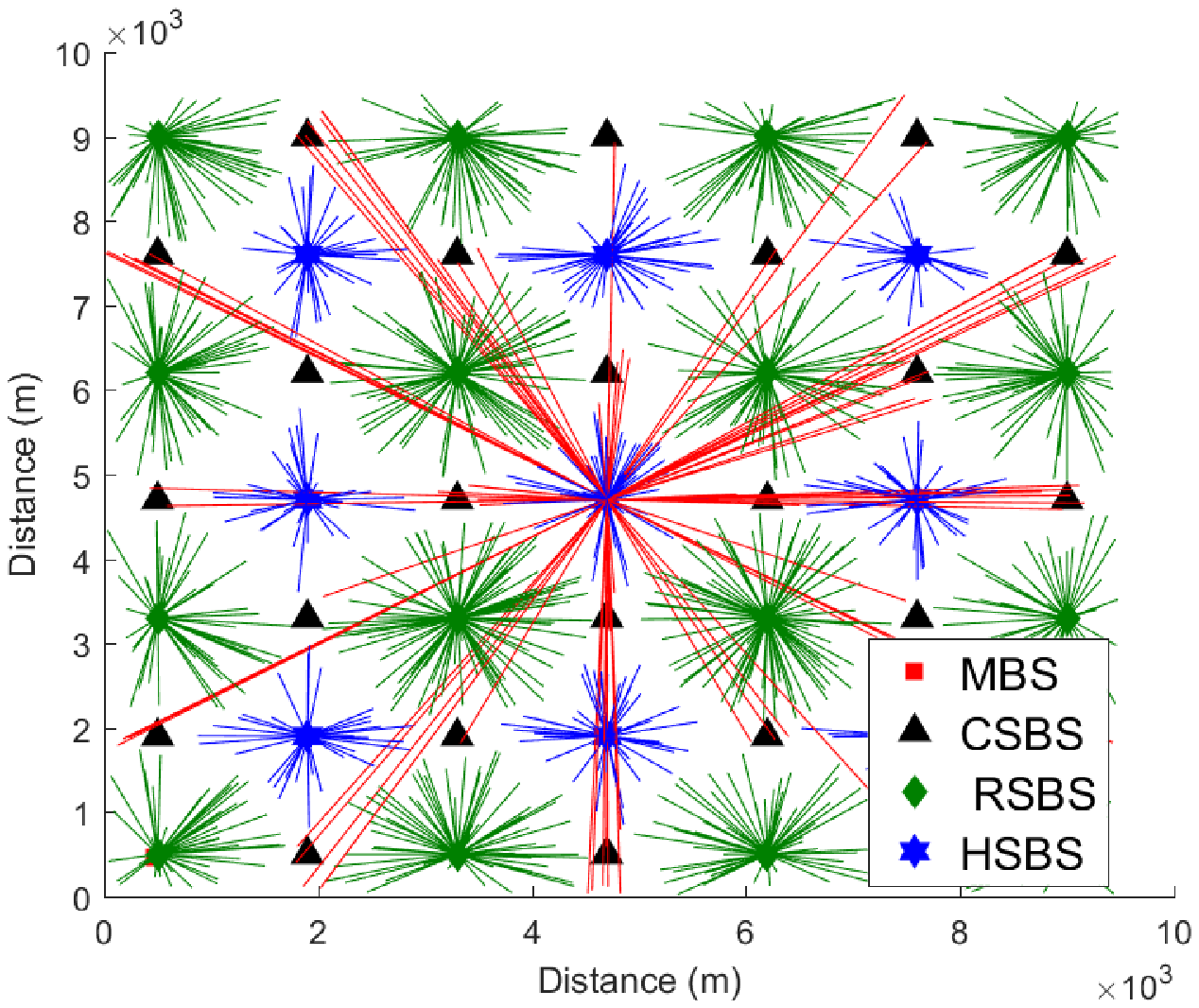}\subcaption{\small {RSBSs, HSBSs \& MBS serve users while CSBSs are in sleep mode.}}\label{a}\par
		\includegraphics[width=1.9in]{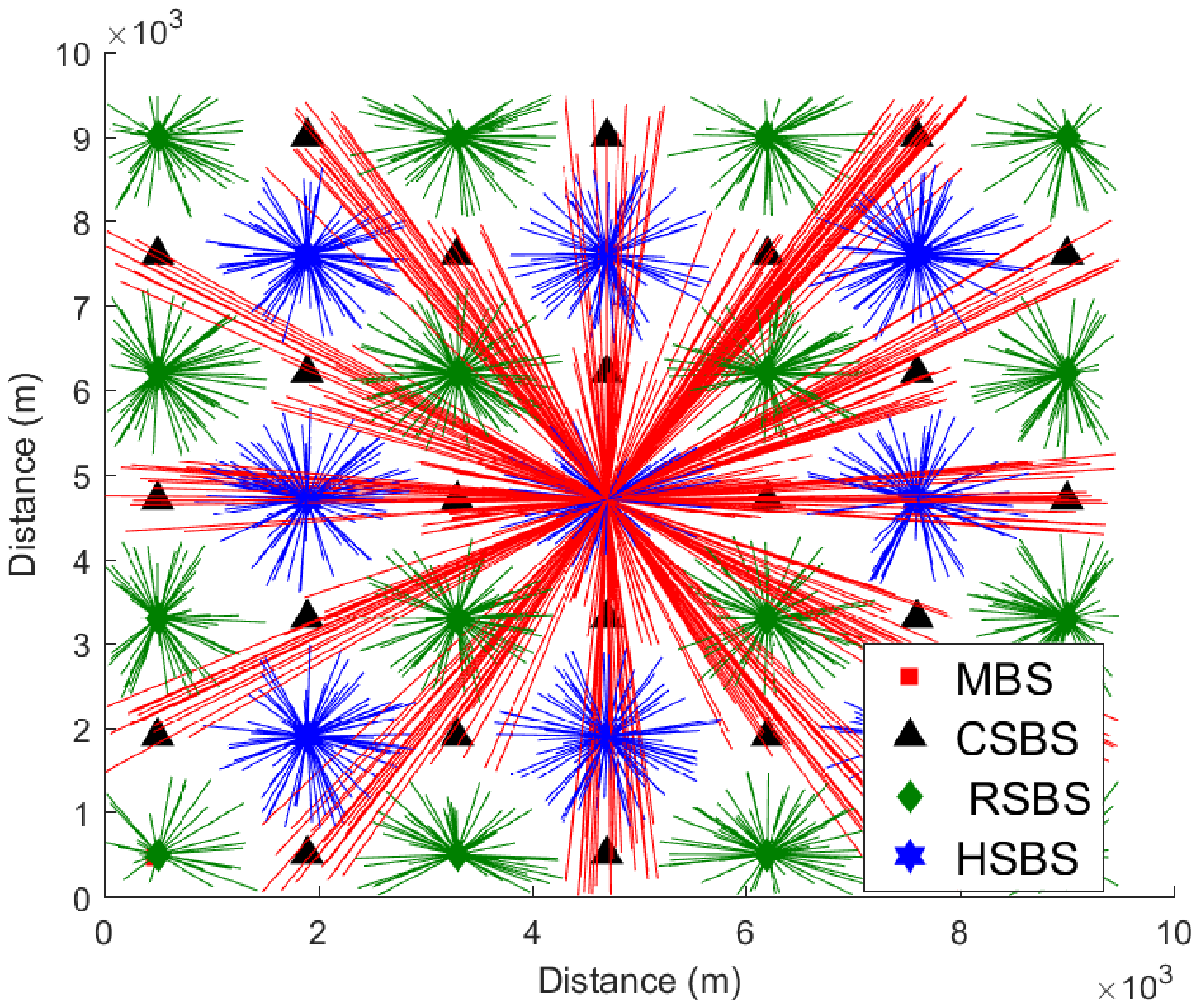}\subcaption{\small {RSBSs reduce and HSBSs increses transmit power to keep CSBSs in sleep mode.}}\label{b}\par
		\includegraphics[width=1.9in]{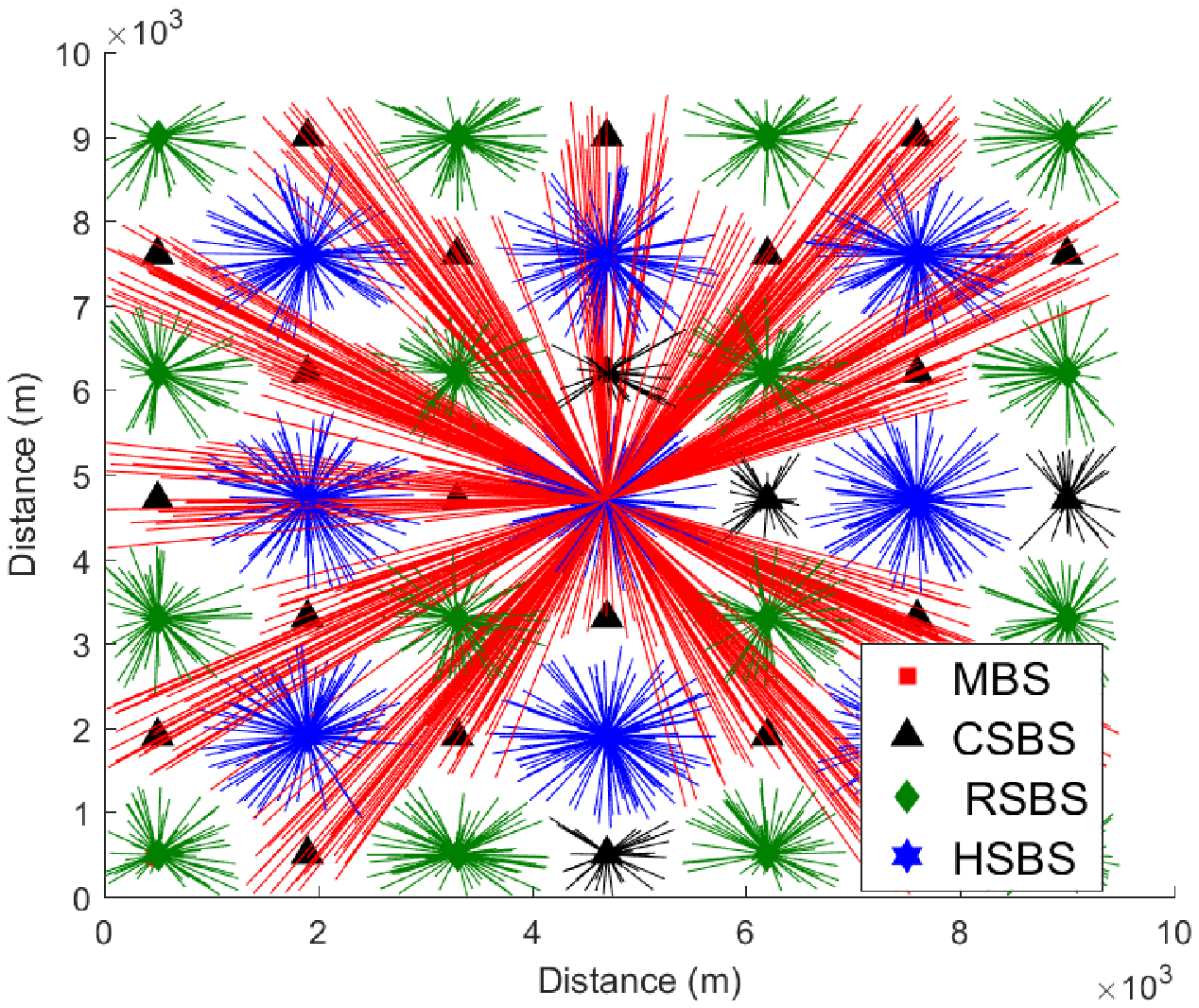}\subcaption{\small {RSBSs reduce power, HSBSs are at max transmit power value \& very few CSBSs are in active mode.}}\label{c}\par
		\includegraphics[width=1.9in]{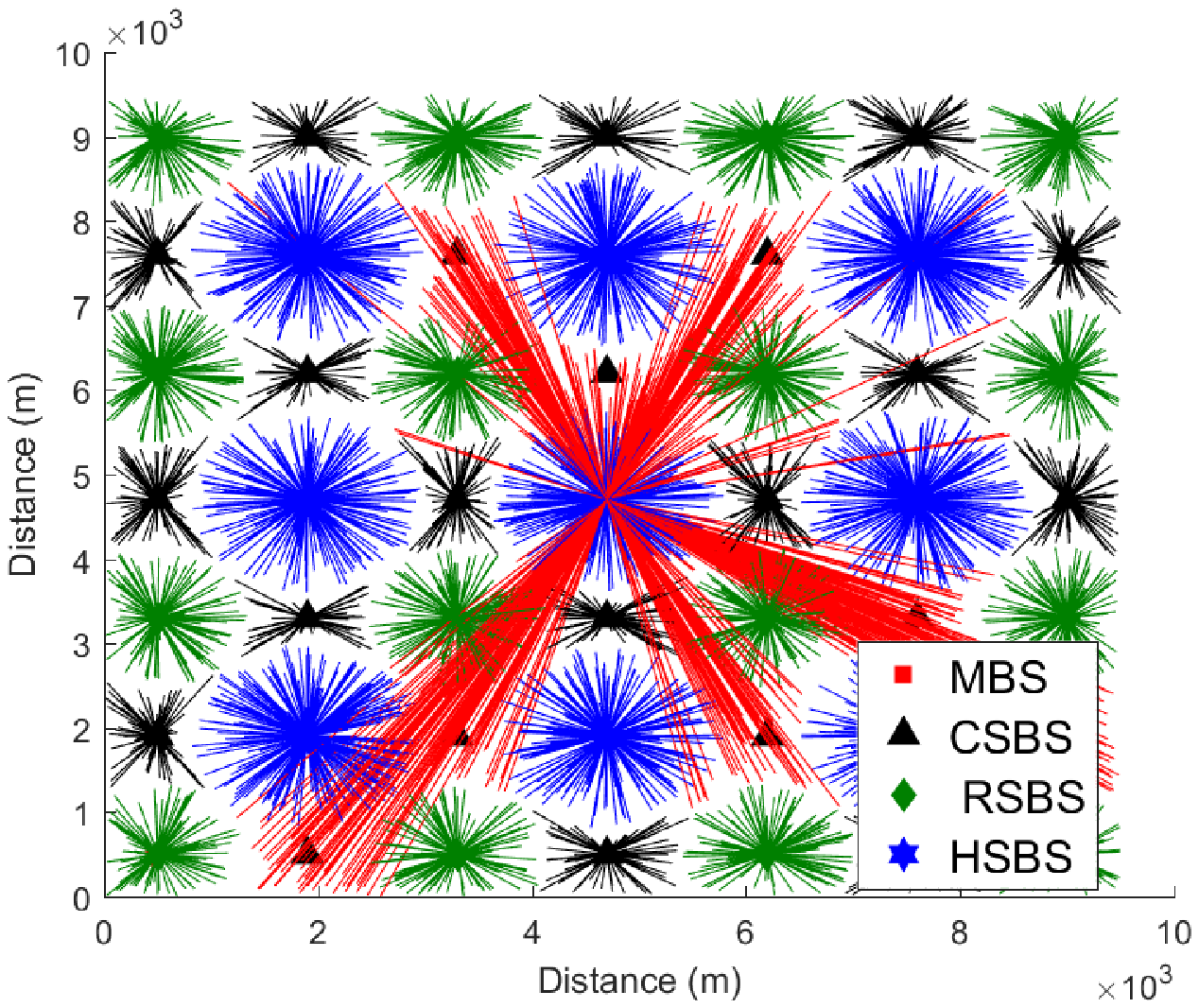}\subcaption{\small {Activated CSBSs increase transmit power while MBS serves at full capacity keeping few CSBSs in sleep mode.}}\label{d}\par
	\end{multicols}
	\label{user}
	\caption{Examples of associations in HCN with varying user densities (increasing from (a) to (d))}
\vspace{-0.4cm}
\end{figure*}

In the proposed scheme, each EH-SBS changes its cell size to balance the load with the harvested energy. If the number of users is small, only EH-SBSs and MBS provide services to the users by moving the CSBSs in sleep mode, as illustrated in Fig.~\ref{a}. As the density of the users increases, a small number of RSBSs and MBS cannot provide services to all the users. Hence RSBSs offload the users by reducing the cell size and the HSBSs increase the coverage area to provide services to these offloaded users by utilizing both harvested energy and on-grid power. This scenario is illustrated in Fig.~\ref{b}. A further increase in the number of users results in coverage area reduction of RSBSs. When HSBS cell size reaches to its maximum, some of the CSBSs are switched from the sleep mode to active mode in order to fill the coverage holes of the system as represented in Fig.~\ref{c} and~\ref{d}.
	
We consider three different association schemes for our network setup. \textit{Nearest BS Scheme} having BSs always in active mode and the users are associated with the nearest SBSs. \textit{Joint Association Scheme} where a user, uses Algorithm 1 to associate with the BS and an SBS dynamically activates/deactivates and changes its coverage area to maximize the energy efficiency (represented in Algorithm 1 (step 1-27). \textit{Proposed Joint Association Scheme} uses complete Algorithm 1 (including lines 28-31) for user association and SBS dynamic activation and deactivation. In this scheme, MBS operates at its full capacity by moving some of the CSBSs in sleep mode.\footnote{Note that the only difference between Joint Association scheme and the Proposed Joint Association scheme is the fact that MBS can identify idle SBSs and deactivate them to save more energy at the cost of more power consumption from the MBS.}
\subsection{On-grid Power Reduction}
\begin{figure}[!t]
	\centering
	\includegraphics[width=3.3in]{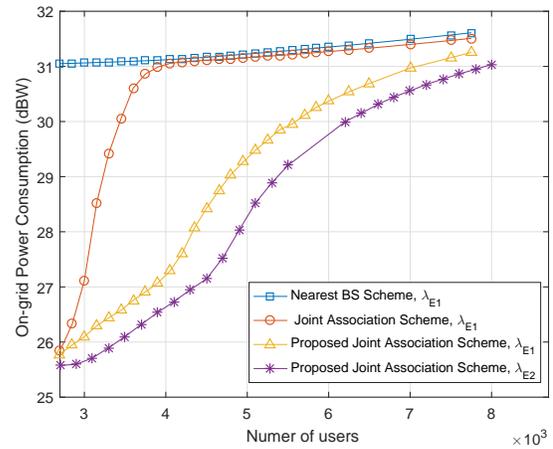}
	\caption{\small Comparison of power consumption with varying traffic load and energy arrival rate for various offloading schemes.}
\vspace{-0.4cm}
	\label{pc}
\end{figure}
 On-grid power consumption of our proposed scheme is compared in Fig.~\ref{pc} which shows less power consumption as compared to other schemes. It can be seen for energy arrival rate $\lambda_{E1} $ = 44 J/s, the joint scheme achieves high power saving  when the user density is low. This is because the constant power consumption is reduced by turning the CSBSs into sleep mode. Moreover, the proposed joint scheme having arrival rate $\lambda_{E1}$ can achieve the best performance for low and moderate number of users by moving more CSBSs in sleep mode while the MBS operates at its full capacity. Compared with nearest BS scheme, our scheme saves 72\% on-grid power for small number of users. It can be seen that the proposed scheme achieves high power saving when the energy arrival rate is high ($\lambda_{E2}$ =45 J/s).
\subsection{Network Rate}
\begin{figure}[!t]
	\centering
	\includegraphics[width=3.3in]{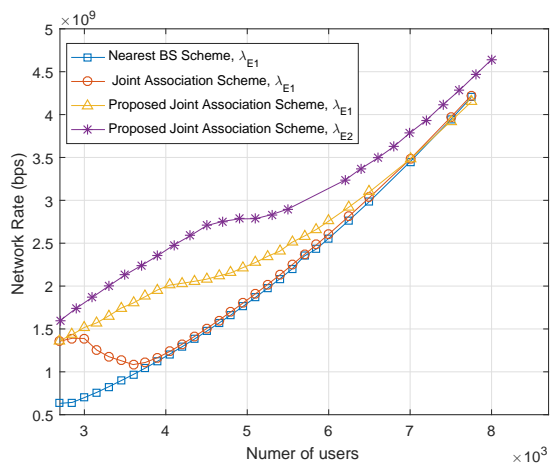}
	\caption{\small Comparison of network throughput with varying traffic load and energy arrival arrival rate for different traffic offloading schemes.}
\vspace{-0.4cm}
	\label{t}
\end{figure}

Fig.~\ref{t} shows the network rate with varying user density for different traffic offloading schemes. When the proposed scheme is used in our system model with energy arrival rate $\lambda_{E1}$, the network rate is higher than the nearest BS  scheme, especially when the user density is low. This is due to small number of SBSs in active mode which causes less interference with other BSs. As density of users increases, HSBSs increase their cell size by utilizing both energy harvesting and on-grid power, if the energy from the renewable source alone is not enough. To balance the load with harvested energy, RSBSs reduce their cell size and more CSBSs are switched into active mode, in order to give more system capacity. Both Joint and the Proposed schemes change the coverage area of SBSs in order to reduce the interference with other cells.

\subsection{Energy Efficiency}
\begin{figure}[!t]
	\centering
	\includegraphics[width=3.3in]{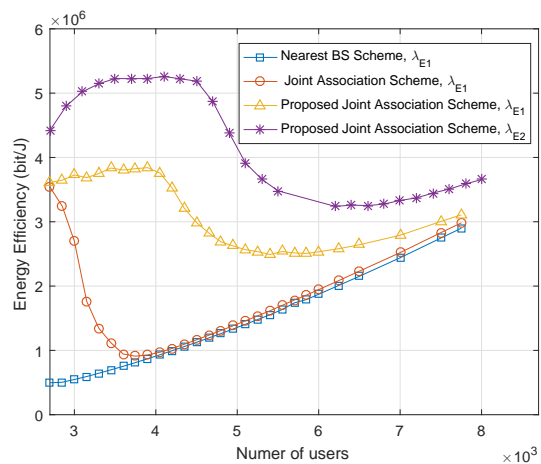}
	\caption{\small Comparison of energy efficiency with varying traffic load for different traffic offloading schemes.}
\vspace{-0.4cm}
	\label{EE}
\end{figure}

We compare network energy efficiencies of various schemes with energy arrival rate $\lambda_{E1}$ = 44 J/s and $\lambda_{E2}$ = 45 J/s in Fig.~\ref{EE}. For low user density, energy efficiency for the joint scheme is higher than the nearest BS scheme due to less number of SBSs in active mode, which consumes less amount of on-grid power. For high user density, energy efficiency of the proposed scheme decreases as small number of CSBSs switch into active mode. The proposed joint association scheme keep more SBSs in sleep mode and changes cell size to increase system capacity by minimizing the interference. Therefore, energy efficiency for both, low and medium traffic, is higher than the nearest BS scheme. It can be seen that the proposed scheme achieves higher energy efficiency when the energy arrival rate is high, especially for the moderate number of users.

\section{Conclusion}
We address joint user association and cell activation framework in a heterogenous network with the aim to minimize on-grid power consumption and increase system energy efficiency. In the proposed scheme, conventional and energy harvesting base stations optimize their coverage area and activation states to minimize the interference and balance the network load with explicit knowledge of energy states of energy harvesting small cell base stations. System simulations show that the proposed joint scheme outperforms other schemes in terms of minimizing on-grid power consumption and maximizing system capacity and energy efficiency. At low load, smartly performed deactivation of small cells and macro cell coverage extension actually helps to reduce system power consumption due to reduction in interference and elimination of constant part of power consumption in the base stations.

\bibliographystyle{IEEEtran}

\bibliography{biblo}

\begin{thebibliography}{10}
\providecommand{\url}[1]{#1}
\csname url@samestyle\endcsname
\providecommand{\newblock}{\relax}
\providecommand{\bibinfo}[2]{#2}
\providecommand{\BIBentrySTDinterwordspacing}{\spaceskip=0pt\relax}
\providecommand{\BIBentryALTinterwordstretchfactor}{4}
\providecommand{\BIBentryALTinterwordspacing}{\spaceskip=\fontdimen2\font plus
\BIBentryALTinterwordstretchfactor\fontdimen3\font minus
  \fontdimen4\font\relax}
\providecommand{\BIBforeignlanguage}[2]{{%
\expandafter\ifx\csname l@#1\endcsname\relax
\typeout{** WARNING: IEEEtran.bst: No hyphenation pattern has been}%
\typeout{** loaded for the language `#1'. Using the pattern for}%
\typeout{** the default language instead.}%
\else
\language=\csname l@#1\endcsname
\fi
#2}}
\providecommand{\BIBdecl}{\relax}
\BIBdecl

\bibitem{piro2013hetnets}
G.~Piro, M.~Miozzo, G.~Forte, N.~Baldo, L.~A. Grieco, G.~Boggia, and P.~Dini,
  ``Hetnets powered by renewable energy sources: Sustainable next-generation
  cellular networks,'' \emph{IEEE Internet Computing}, vol.~17, no.~1, pp.
  32--39, 2013.

\bibitem{zhang2015many}
S.~Zhang, J.~Gong, S.~Zhou, and Z.~Niu, ``How many small cells can be turned
  off via vertical offloading under a separation architecture?'' \emph{IEEE
  Transactions on Wireless Communications}, vol.~14, no.~10, pp. 5440--5453,
  2015.

\bibitem{zhang2014energy}
S.~Zhang, J.~Wu, J.~Gong, S.~Zhou, and Z.~Niu, ``Energy-optimal probabilistic
  base station sleeping under a separation network architecture,'' in
  \emph{Global Communications Conference (GLOBECOM)}, 2014.

\bibitem{gong2014base}
J.~Gong, J.~S. Thompson, S.~Zhou, and Z.~Niu, ``Base station sleeping and
  resource allocation in renewable energy powered cellular networks,''
  \emph{IEEE Trans. on Communications}, vol.~62, no.~11, pp. 3801--3813, 2014.

\bibitem{feng2014energy}
J.~Feng, M.~Yinxia, P.~Wang, X.~Zhang, and W.~Wang, ``Energy-aware resource
  allocation with energy harvesting in heterogeneous wireless network,'' in
  \emph{International Symposium on Wireless Communications Systems (ISWCS)},
  2014, pp. 766--770.

\bibitem{kim2013joint}
S.~Kim, S.~Choi, and B.~G. Lee, ``A joint algorithm for base station operation
  and user association in heterogeneous networks,'' \emph{IEEE Communications
  Letters}, vol.~17, no.~8, pp. 1552--1555, 2013.

\bibitem{Energy-Awaretraffic}
S.~Zhang, N.~Zhang, S.~Zhou, J.~Gong, Z.~Niu, and X.~Shen, ``Energy-aware
  traffic offloading for green heterogeneous networks,'' \emph{IEEE Journal on
  Selected Areas in Communications}, vol.~34, no.~5, pp. 1116--1129, 2016.

\bibitem{ye2013user}
Q.~Ye, B.~Rong, Y.~Chen, M.~Al-Shalash, C.~Caramanis, and J.~G. Andrews, ``User
  association for load balancing in heterogeneous cellular networks,''
  \emph{IEEE Transactions on Wireless Communications}, vol.~12, no.~6, pp.
  2706--2716, 2013.

\bibitem{jo2012heterogeneous}
H.-S. Jo, Y.~J. Sang, P.~Xia, and J.~G. Andrews, ``Heterogeneous cellular
  networks with flexible cell association: A comprehensive downlink sinr
  analysis,'' \emph{IEEE Transactions on Wireless Communications}, vol.~11,
  no.~10, pp. 3484--3495, 2012.

\bibitem{muhammad2017cell}
F.~Muhammad, Z.~H. Abbas, and F.~Y. Li, ``Cell association with load balancing
  in nonuniform heterogeneous cellular networks: Coverage probability and rate
  analysis,'' \emph{IEEE Transactions on Vehicular Technology}, vol.~66, no.~6,
  pp. 5241--5255, 2017.

\bibitem{krikidis2013harvest}
I.~Krikidis, G.~Zheng, and B.~Ottersten, ``Harvest-use cooperative networks
  with half/full-duplex relaying,'' in \emph{Wireless Communications and
  Networking Conference (WCNC)}, 2013, pp. 4256--4260.

\bibitem{auer2010d2}
Auer \emph{et~al.}, ``Energy efficiency analysis of the reference systems,
  areas of improvements and target breakdown,'' 2010, deliverable D2.3, EARTH
  project.

\bibitem{Personalcell_Traffic_offloading}
P.-S. Yu, J.~Lee, T.~Q. Quek, and Y.-W.~P. Hong, ``Traffic offloading in
  heterogeneous networks with energy harvesting personal cells-network
  throughput and energy efficiency,'' \emph{IEEE Transactions on Wireless
  Communications}, vol.~15, no.~2, pp. 1146--1161, 2016.

\bibitem{tombaz2011impact}
S.~Tombaz, P.~Monti, K.~Wang, A.~Vastberg, M.~Forzati, and J.~Zander, ``Impact
  of backhauling power consumption on the deployment of heterogeneous mobile
  networks,'' in \emph{IEEE Global Telecommunications Conference (GLOBECOM)},
  Dec 2011, pp. 1--5.

\end{thebibliography}
%
%
%
%

\end{document}